\documentclass[12pt,apjpt4]{aastex}

\shorttitle{An Infrared Imaging Study of IRAS16594$-$4656}

\shortauthors{Volk, Hrivnak, Su, \& Kwok}

\begin{document}


\title{An Infrared Imaging Study of the Bipolar Proto-Planetary Nebula IRAS 
16594$-$4656\altaffilmark{1}}


\author{Kevin Volk}
\affil{Gemini Observatory, 670 N.~A'ohoku Place, Hilo, HI 96720; 
kvolk@gemini.edu}

\author{Bruce J. Hrivnak}
\affil{Department of Physics and Astronomy, Valparaiso University,
Valparaiso, IN 46383; bruce.hrivnak@valpo.edu}

\author{Kate Y.~L.~Su}
\affil{Steward Observatory, University of Arizona, Tucson, AZ 85721; 
ksu@as.arizona.edu}

\and

\author{Sun Kwok}
\affil{Department of Physics, The University 
of Hong Kong, Hong Kong, China; and Department of Physics and 
Astronomy, University of Calgary, Calgary, Alberta, Canada; sunkwok@hku.hk}


\altaffiltext{1}{The paper is based on observations obtained at
the Gemini Observatory.  The Gemini Observatory is operated by the Association
of Universities for Research in Astronomy, Inc., under a
cooperative agreement with the NSF on behalf of the Gemini
partnership: the National Science Foundation (United States), the
Particle Physics and Astronomy Research Council (United Kingdom),
the National Research Council (Canada), CONICYT (Chile), the
Australian Research Council (Australia), CNPq (Brazil) and CONICET
(Argentina).}


\begin{abstract}

High-resolution mid-infrared images have been obtained in N-band and 
Q-band for the proto-planetary nebula IRAS 16594$-$4656.  A bright 
equatorial torus and a pair of bipolar lobes can clearly be seen in the 
infrared images.  The torus appears thinner at the center than at the 
edges, suggesting that it is viewed nearly edge-on.  The infrared lobes 
correspond to the brightest lobes of the reflection nebula seen in the 
Hubble Space Telescope ({\it HST}) optical image, but with no sign of the 
point-symmetric structure seen in the visible image.  The lobe structure 
shows a close correspondence with a molecular hydrogen map obtained 
with {\it HST}, suggesting that the dust emission in the lobes traces the 
distribution of the shocked gas.  The shape of the bipolar lobes shows 
clearly that the fast outflow is still confined by the remnant 
circumstellar envelope of the progenitor asymptotic giant branch 
(AGB) star.  However, the non-detection of the dust outside of the 
lobes suggests that the temperature of the dust in the AGB envelope 
is too low for it to be detected at 20 $\mu$m.

\end{abstract}




\section{INTRODUCTION}

Proto-planetary nebulae (PPNe) are the long-sought-after missing link 
between the end of the asymptotic giant branch (AGB) phase and the 
beginning of planetary nebula phase of stellar evolution.  After the 
{\it Infrared Astronomical Satellite} ({\it IRAS}) mission, a number of 
objects were proposed as candidate PPNe based on their infrared colors 
and other spectral properties. These are typically stars of G to B spectral 
type with significant infrared excesses due to the remnant circumstellar 
dust shell ejected in the AGB phase.  Of particular interest among these 
candidates are a number of carbon-rich objects whose abundances show a 
strong enhancement of s-process elements, as expected from the dredge-up 
of material in thermal pulses during the AGB evolution \citep{kwok93, 
vanwinckel03}.  For other candidates there is some possibility of confusion 
with massive supergiants, but those in this carbon-rich group are almost 
certainly bona-fide PPNe.

One of these objects is IRAS 16594$-$4656.  It is a bright 
mid-infrared source which has typical colors of a PPN \citep{volk89}.  
Optically it is associated with a southern emission-line star. It was 
found to be of spectral type B7 with V magnitude 14.6, subject to about 
7.5 magnitudes of visual extinction \citep{steene00}.  It was not clear 
from the original {\it IRAS} spectral observations whether the object was 
oxygen-rich or carbon-rich, but subsequent {\it Infrared Space Observatory} 
({\it ISO}) spectral observations showed that it has carbon-based dust 
features, including the 21 $\mu$m feature \citep{garcia99}.  Optical images 
obtained with the {\it Hubble Space Telescope} showed that 
the star has a surrounding reflection nebula with a complex structure 
\citep{hrivnak99}.  The relative faintness of the reflection nebula 
compared to the star, together with the optical morphology, led 
\citet{hrivnak99} to conclude that the nebula is intrinsically bipolar 
(or multi-polar) viewed at an intermediate angle to the bipolar axis.  
Both optical and near-infrared spectral observations show emission lines, 
which are thought to be shock-excited rather than radiatively excited 
by the star \citep{garcia99,steene03}.  No radio emission has 
been detected from IRAS 16594$-$4656 \citep{steene93}.

The distance to this object is uncertain.  The dust shell model of 
\citet{hrivnak00} suggests a distance of 2.6 kpc if the total luminosity 
of the star is 10$^4$ $L_\odot$.  A slightly smaller estimate of (2.2$\pm$0.4) 
kpc is given by \citet{steene03} for the same assumed total luminosity.  
Most of the luminosity is being emitted in the infrared where extinction 
effects are smaller than in the optical, so these estimates are not 
strongly affected by the non-spherical morphology of the dust shell.

The optical morphology of the nebula in IRAS 16594$-$4656 appears complex, 
with what appear to be pairs of symmetric structures at multiple position 
angles,  for which it was named the ``Water Lily Nebula'' \citep{hrivnak99}.
The optical images also showed several concentric arcs centered on the 
star \citep{hrivnak01}.  However, near-infrared observations with the 
{\it NICMOS} instrument on {\it HST} showed a somewhat simpler morphology, 
although it was not clear whether this was simply an effect of reduced 
dynamic range compared to the optical observations \citep{su03}.  An initial 
N-band observation of IRAS 16594$-$4656 with the TIMMI2 camera on the  
ESO 3.6m telescope failed to resolve the dust shell \citep{steene00},  
while more recent TIMMI2 observations did marginally resolve the structure 
\citep{garcia04}.

In this paper we present new, higher sensitivity and higher angular resolution 
mid-infrared images of IRAS 16594$-$4656 which resolve the dust shell in 
thermal emission.  We present the observations in section 2.  We then derive 
a dust color temperature map in section 3, and compare the morphology 
as observed in the mid-infrared to that seen at other wavelengths in 
section 4. We give a brief discussion of the results in the final section.

\section{OBSERVATIONS}

The observations reported here were obtained with the T-ReCS instrument on 
Gemini South under program GS-2004A-Q-56.  The sky and telescope background 
were removed by chopping and nodding during the observations.  Images of 
IRAS 16594$-$4656 were obtained in three filters.  In the N-band window: 
360 second (total on-source exposure times) images with the ``Si-5 11.66um'' 
filter on 2004 March 11 and with the ``Si-6 12.33um'' filter on 2004 May 8; 
and in Q-band window: a 600 second image with the ``Qa 18.30um'' filter on 
2004 May 10.  These filters will be referred to as the ``Si5'', ``Si6'', 
and ``Qa'' filters, respectively.  All these filters have widths $\Delta\lambda \sim 1 
\mu{\rm m}$.  Information about these filters, including the filter profiles, 
can be found on the Gemini Observatory public WWW pages (see 
{\it http://www.gemini.edu/sciops/instruments/miri/T-ReCSFilters.html}).  
On each night, standard star observations for flux calibration were carried 
out immediately after the observation of the science target.  These stars 
were HD 123139 on March 11, HD 169916 on May 8, and HD 175775 on May 10.  
All of these stars are included among the mid-infrared spectrophotometric 
standards of \citet{cohen99}.  Comparisons of the point-spread functions 
(PSFs) of these stars with observations of $\alpha$ Cen A/B or $\alpha$ CMa 
on various nights from December 2003 through May 2004 indicate that these 
stars are suitable as PSF references as well as spectral standards.

The N-band observations were made under good conditions, as judged from the 
level of sky cancellation obtained while chopping.  The Q-band observations 
were made under marginal conditions; however IRAS 16594$-$4656 is a 
very bright target at 18 $\mu$m and it was detected with good signal-to-noise 
ratio despite the less than ideal conditions.  The image quality was good as 
estimated from the standard stars.  The full width at half maximum (FWHM) was 
0$\farcs$37 for the Si5 filter image, 0$\farcs$39 for the Si6 image, and 
0$\farcs$60 for the Qa image.  These correspond to Strehl values of about 
0.6, fairly typical of better seeing conditions for the N-band filters but 
somewhat lower than usual for the Qa filter.

The raw images on and off of the target were subtracted and then summed to 
produce raw images of IRAS 16594$-$4656 in the three filters. Flux calibration
of the images was done in two different ways.  First, the standard star 
observations were used to find the conversion from in-band counts to Jy 
using the assumed spectral energy distribution from \citet{cohen99} 
integrated over the filter profiles, and then these scaling factors were 
applied to the images of IRAS 16594$-$4656.  The pixel by pixel brightnesses 
so obtained were then converted to Jy/square arc-second using the pixel size 
of T-ReCS.  Second, the estimated filter flux densities were generated from 
the {\it ISO} spectrum of IRAS 16594$-$4656 in the same way as was used to 
get the expected flux densities in Jy for the standard stars.  These values 
were also used to convert from counts to Jy/square arc-second in the images.  
It was found that for the Si5 filter these two methods agreed within 
less than 1\%.  For the other filters the agreement was poorer.  Using the 
{\it ISO} spectrum for the Si6 filter estimate gave a value 22\% higher 
than that from the standard star.  This discrepancy is too large to be due 
to an atmospheric extinction effect, judging from some estimates of the 
extinction coefficient in this filter made on other nights.  It probably 
indicates that the sky conditions were not uniform in the directions to 
the standard star and to our target, as in other observations we have 
obtained with T-ReCS the inter-comparison with {\it ISO} spectra gives 2\% 
agreement for filters in the N-band window.  The Qa brightness 
calculated from the standard star came out 11\% lower than that 
calculated from the {\it ISO} spectrum.  This is probably within the 
uncertainties caused by the variable sky conditions.  There is also 
a color effect due to differences in spectral shape between 
IRAS 16594$-$4656 and the standard stars, but since the filters are 
relatively narrow this is a small correction and it was neglected.

In what follows, we have chosen to use the {\it ISO} spectrum as the basis for 
creating surface brightness images, since this minimizes the effect of
variable sky conditions.  This assumes that the {\it ISO} spectrum 
gives the correct absolutely calibrated total brightness and the T-ReCS 
observations give the correct relative brightness distribution, or 
equivalently that all the atmospheric effects are uniform over the 
small T-ReCS field of view.  Figure \ref{fig1} shows the three 
flux-calibrated images in Jy/square arc-second.  The region shown 
in the figure is $7\farcs 2\times7\farcs 2$ and contains all the 
detected emission from IRAS 16594$-$4656.  The total flux density 
for IRAS 16594$-$4656 was calculated to be 44.0, 56.9, and 177 Jy for 
the Si5, Si6, and Qa filters, respectively.

  In all three cases, the circumstellar shell of IRAS 16594$-$4656 appears 
as a bipolar nebula of dimension about 4$\farcs$5 $\times$ 2$\farcs$25, with a 
bright central region orientated roughly north-south and two lobes extending 
east and west.  There is a clear difference in size between the east and west 
lobes.  The east lobe is about 20\% smaller than the west lobe both in width 
and maximum detected radius from the star in these images.  The optical depth 
at these wavelengths, especially in Q-band, must be small; thus this size 
difference must be caused by either a physical size difference between the 
two lobes or distinctly different projection angles for the two lobes.

The central bright region of the mid-infrared images appears to be some type 
of thin ``equatorial'' torus, perpendicular to the axis of the two lobes.  
The northern end of the this structure is brighter than the southern end in 
all the images, but the ratio is much closer to 1:1 in the Qa image.  This 
shows that the dust in the torus is cooler in the southern region than in the 
northern region.  The sharp edge of the torus in the north is particularly 
striking, as the images are very bright there but there is a sudden edge 
beyond which no emission is observed along the line of the torus.  

\section{IMAGE ANALYSIS}

\subsection{Image Deconvolution}

Lucy deconvolution of the raw images was carried out using the 
standard star observations as PSF templates.  This was done using 
the stsdas.analysis.restore.lucy task in the STSCI reduction package 
under IRAF version 2.12a.  A 61 by 61 pixel box was used to define the 
PSF.  Pixels outside this PSF box but within a 181 by 181 pixel 
box centered on the star were used to derive the background level, which 
was subtracted from the stellar profile.

It was found that most of the improvement in the image resolution was 
obtained in the first few Lucy iterations, after which there was no 
significant change in the derived structure, so the deconvolution was 
stopped after 20 iterations.  The deconvolved images were then 
re-smoothed with a gaussian of FWHM 0$\farcs$1, slightly 
larger than the original pixel size.  The resulting images are 
sharpened by about a factor of 3 in PSF width compared to the original images. 

Figure \ref{fig2} shows these sharpened images.  The two N-band images have 
almost identical structure.  The sharpened Q-band image has the 
same general morphology as the N-band images but both lobes are 
seen to be smaller by about 0.4 arc-seconds than in the N-band images.  
However, we believe that this size difference is not real.  
Comparison of the raw images in the Qa and N-band filters shows that 
the emission region is just slightly larger for the Qa image than for 
the Si5  and Si6 filters.  This indicates that the Lucy deconvolution 
for the lobes introduced a small artifact into the image.  As the Strehl 
value for the standard star in the Qa filter was lower than usual, it is 
possible that the seeing changed between the target observation 
and the standard star observation.  There was some indication of variable 
Q-band conditions during these observations.  If the seeing did get worse 
for the standard star observation, that would explain the decrease in lobe 
size for the deconvolved image compared to the raw image, although the 
magnitude of the decrease is larger than one would expect based upon the 
FWHM value for the standard star Qa filter image.

The structure of the brightest regions is similar for all three filters.  
The bright part suggests a torus of some sort, and it appears to be thinner 
at the center than at the edges.  This suggests that the torus is seen nearly
edge on.  If it were oriented at some intermediate angle to the 
plane of the sky, as was earlier asserted based upon the visible images, 
then one would expect the torus to appear as a small ellipse in these 
mid-infrared images.  This is clearly not seen.  However, if we are indeed 
viewing the torus edge-on then the visible and mid-infrared images 
suggest it to be quite asymmetric with position.

In each panel of Figure \ref{fig2}, the estimated star position is marked 
with a small black dot.  This position was found by cross-comparing 
the T-ReCS Si5 image with an {\it HST NICMOS} image taken in the 
narrow-band filter centered on the H$_2$ 2.12 $\mu$m line \citep{hrivnak04}.  
The estimated stellar position is very close to the geometrical center 
of the bright ``bar'' in the dust emission.  It is located in a region of 
relatively low brightness in the three filters, especially in the 
Si5 and Si6 filters.  This probably means that in the optical images 
we are seeing the star through some type of hole in the toroid, 
where little dust is present.

\subsection{Dust Color Temperature Maps}

From the surface brightness images in two filters it is possible to construct 
a ratio map, and if one assumes that the surface brightness is due to 
optically-thin thermal emission from dust grains of a known type, then the 
surface brightness ratio can be converted into color temperatures between 
the two wavelengths.  Denoting the surface brightness in Jy/square arc-second 
by $S(\nu)$, the color temperature, $T_c$, is defined by solving 
$$ 
{{S(\nu_1)}\over{S(\nu_2)}} = {{\tau_{\nu_1} B_\nu(T_c,\nu_1)}
\over{\tau_{\nu_2} B_\nu(T_c,\nu_2)}}
$$
for each brightness ratio value in the image.  The optical depth values 
$\tau_\nu$ for the two filters are assumed to be proportional to the 
absorption cross-sections, so we can replace the optical depths with the 
actual Q${\rm abs}$ values for this calculation.  The scattering component 
of the dust extinction is expected to be small at these long wavelengths.  
The ratio of $\tau_\nu$ values is a constant as long as the dust grain 
properties are uniform in the dust shell, so then the equation gives a 
one-to-one transformation between surface brightness ratio and $T_c$.

We have used the Si5 and Qa images to produce a brightness ratio map, after 
convolving the Si5 image with a gaussian profile of FWHM of 0$\farcs$195 to 
match the angular resolution of the Qa image.  Regions of the two images 
which had ``low'' brightness values, taken to be less than 0.1\% of the 
respective peak values, were masked out of the ratio image.  The ratio 
image was then transformed to dust color temperature, assuming that the 
dust grains are 0.1 $\mu$m amorphous carbon grains with the opacity 
function for AC type 2 grains from \citet{rouleau91}.  These grain 
properties were the basis of the spectral model for IRAS 16594$-$4656 
presented by \citet{hrivnak00}.  For another assumed dust grain size 
or type the dust color temperature values would change, but the relative 
variations over the image should remain the same.  While the $T_c$ values 
do not indicate the physical temperatures of the dust grains, since they 
are some type of average along the line of sight for each pixel, they 
do indicate the global dust temperature variations in the circumstellar 
shell as long as the dust grains are not drastically different than 
assumed for the calculation.  The $T_c$ map is shown in the lower 
right panel of Figure \ref{fig1}.  The $T_c$ values are confined to a 
relatively narrow range.  The bright part of the dust shell has a 
range of $T_c$ from about 140 K to 160 K.  It was found that the $T_c$ map was 
much the same whether or not the Si5 image was convolved to the resolution of
the Qa image.

The $T_c$ map shows that the region of highest dust color temperature is 
in the north part of the central bright region, while the color temperature 
is much lower directly to the south on the other side of the stellar position. 
There is also a region of high dust color temperature in the southern wall 
of the west lobe.

There is a cap at the end of the east lobe that is seen in the color 
temperature map, which is also apparent in the Si5 image.  This cap 
does not appear to be associated with a higher dust color temperature 
than elsewhere in lobe, which suggests that the dust optical depth is 
higher here than for other positions in the lobe.  It is possible that the 
east lobe is smaller than the west lobe because its expansion is impeded by 
external material, and that the cap represents a boundary between the lobe 
and the external medium.

\section{DISCUSSION and CONCLUSIONS}

Figure \ref{fig3} shows the deconvolved Si5 image with overlaid contour 
plots in the optical I-band (0.8 $\mu$m; Su, Hrivnak, \& Kwok 2001) 
and near-infrared H$_2$ filter images (2.12 $\mu$m; Hrivnak, Kelly, \& Su 
2004).  This allows direct comparison of the morphology of the nebula in 
these different wavelengths.  The H$_2$ image was matched to the Si5 filter 
image since they were immediately seen to have very similar morphologies.  
This was used to obtain a reasonably accurate estimate of the central star 
position in the T-ReCS image.  That position was then used to match the 
optical image to the mid-infrared image.  The Si5 image is plotted in 
absolute surface brightness units, Jy/square arc-second.

As shown in the upper right panel of Figure \ref{fig3}, there is a close match 
of features in the lobes between the {\it NICMOS} H$_2$ filter image and the 
T-ReCS image.  The bright lobe edges are regions of strong H$_2$ emission. 
The cap in the east lobe is clearly visible in this panel, and it also 
corresponds to a region of strong H$_2$ emission.  It is more difficult to 
determine if any of the mid-infrared bright waist structure is also detected 
in the H$_2$ image, because the star is saturated in the H$_2$ image, but it 
does not look as if anything but the edges of the lobes are detected in the 
H$_2$ image.  Since the H$_2$ emission is mainly shock excited 
\citep{steene03}, this raises the possibility that the dust emission from 
these regions are partially shock-excited.  Another possibility 
that could explain why the dust emission region so closely matches the 
shock is that the dust grains may be much smaller downstream from the shock 
than they are before the shock, and that as a result the small grains are 
transiently heated to relatively high temperatures.

Comparison of the T-ReCS images with the {\it HST} optical images 
\citep{hrivnak99} shows that the two lobes seen in the T-ReCS 
images correspond closely to the central brightest part of the reflection 
nebula.  The total size of the optical reflection nebula is 
12.3$^{\prime\prime}$ by 8.8$^{\prime\prime}$, which is much larger 
than the size of the N-band or Q-band images.  The optical image overlay 
in the lower left panel of Figure \ref{fig3} uses logarithmically spaced 
contours ranging from 0.0085\% of the stellar peak brightness up to the 
peak brightness, with each contour at a level 2.7 times the previous contour.  
The optical reflection nebula is at most about 0.75\% of the stellar peak 
brightness.  The lowest contours show what has been suggested to be 
point-symmetric morphology, with three pairs of oppositely directed features 
at position angles of about 40$^\circ$, 57$^\circ$, and 87$^\circ$ 
east of north as can be measured from the I-band image presented in 
\citet{hrivnak99}.  The optical lobes have been suggested to be caused by 
a rapidly precessing, columnated high-speed wind from the star 
\citep{garcia99}.  The T-ReCS image shows just the two lobes at position 
angle 75$^\circ$.  While there seems to be some very faint mid-infrared 
emission detected in the Si5 filter on a size scale roughly three times 
that of the main mid-infrared lobes, over-plotting this with the optical 
image does not show any correspondences with the faint extended optical 
structure.  In particular, none of this emission is seen just outside 
the bright torus either to the north or to the south.  Kinematic 
observations of the individual optical lobes is needed to determine 
if they are distinct structures or not.  If these point symmetric 
features are not distinct kinematic structures, then perhaps there 
are just two lobes but the optical appearance is due to structure 
in the walls of these lobes which make them look more complex.

The optical emission is due to reflection by dust, and so the larger 
size of the optical images compared to the mid-infrared images shows that 
the dust shell is much larger than is obvious from the N-band or Q-band 
images.  The dust outside the central bright region delineated by the 
shocked H$_2$ emission is clearly much colder than that inside the shocks.
At the ends of mid-infrared torus, in particular, there must be a large 
discontinuity in temperature and optical depth so that there is a large 
change in mid-infrared surface brightness but a much smaller change in the 
optical brightness of the scattered light.

These mid-infrared images suggest that the torus is perpendicular 
to the plane of the sky.  This is consistent with recent observations 
of the bipolar lobes, which indicate that they are nearly in the plane 
of the sky.  Unpublished high resolution long-slit spectra in the 
near-infrared \citep{hrivnak06} have been obtained with the Phoenix 
spectrograph on Gemini South.  Analysis of these spectra, which map 
the molecular hydrogen line at 2.12 $\mu$m for cuts at different position 
angles through the star and the lobes, shows similar velocities for the two 
lobes, indicating that they are oriented very close to the plane of the 
sky.  This was also concluded by \citet{ueta05} based upon near-infrared 
polarimetry along with dust shell modeling of the spectral energy distribution 
of the object.  Since the evidence indicates that the nebular axis is oriented 
very close to the plane of the sky, the east lobe must actually be smaller 
than the west lobe.

The T-ReCS images show that the central bright region of the circumstellar 
dust shell is quite different than that of other well-known bipolar nebulae 
IRAS 17150$-$3224, IRAS 17441$-$2411, Roberts 22, and Hen 3-401, in all of 
which the dust emission is strongly peaked at the stellar position.  For IRAS 
16594$-$4656 the mid-infrared brightness is at a minimum near the stellar 
position and is much higher to the north and the south along the central waist.
Possibly the torus is of much lower optical depth in this object than in the 
others, or it is highly asymmetric with a low optical depth along our line of 
sight.  The latter suggestion is consistent with the visibility of the 
central star.  Certainly for IRAS 17150$-$3224 and IRAS 17441$-$2411 the 
spectral energy distribution implies a relatively high optical depth along 
our line of sight to the star, and the star is not seen in visible light.  
The spectral type of the star in IRAS 16594$-$4656 is much earlier than that 
in the latter two objects, so it may simply be more evolved and the torus may 
have had more time to disperse.

Unlike the hour-glass or open lobes observed in most bipolar planetary 
nebulae (e.g. NGC 6302), the bipolar lobes of IRAS 16594$-$4656 as shown 
in Figure \ref{fig2} are closed and resemble the lobes seen in the PPN 
IRAS 17106$-$3046 and the young planetary nebula Hen 2-320.  The morphology 
of the lobes clearly shows that the lobes are confined by the circumstellar 
medium, and the fast collimated outflow which creates the bipolar lobes has 
not yet broken through the stellar wind of the AGB progenitor.  The 
interaction between the fast and slow winds is clearly delineated by dust 
distribution in the lobes.  We also note that there is "bulge" at the tip 
of the western lobe, which suggests that the high-velocity flow is on the 
verge of breaking out.  In contrast to the cap at the tip of the eastern 
lobe (which represents a pile-up at the wind interface), the western lobe 
may represent a slightly more advanced stage of the breakout, and therefore 
explains the difference in sizes between the two lobes.  In a few hundred 
years, we expect that both lobes will open up into butterfly morphology.  
We are therefore witnessing a critical phase of morphological transformation 
of PNs. 

We have successfully detected the bipolar lobes and a central bright waist 
in mid-infrared images of IRAS 16594$-$4656.  The bright waist suggests
that we are seeing a central torus nearly edge-on.  While this is consistent 
with published polarization and unpublished kinematic results, it 
differs from earlier published interpretations of the visible image 
that concluded that the lobes are seen at an intermediate orientation. 
This emphasizes the need for multi-wavelength observations to 
confidently understand the structure of proto-planetary nebulae.
This result may well be applicable to our understanding of other 
bipolar phenomenon such as YSOs and AGNs.




\acknowledgments

B.J.H acknowledges support by the National Science Foundation under 
Grant No.~0407087.  This work was supported in part by grants to 
S.K from the Natural Sciences and Engineering Research Council of 
Canada.






\clearpage

\epsscale{1.0} \plotone{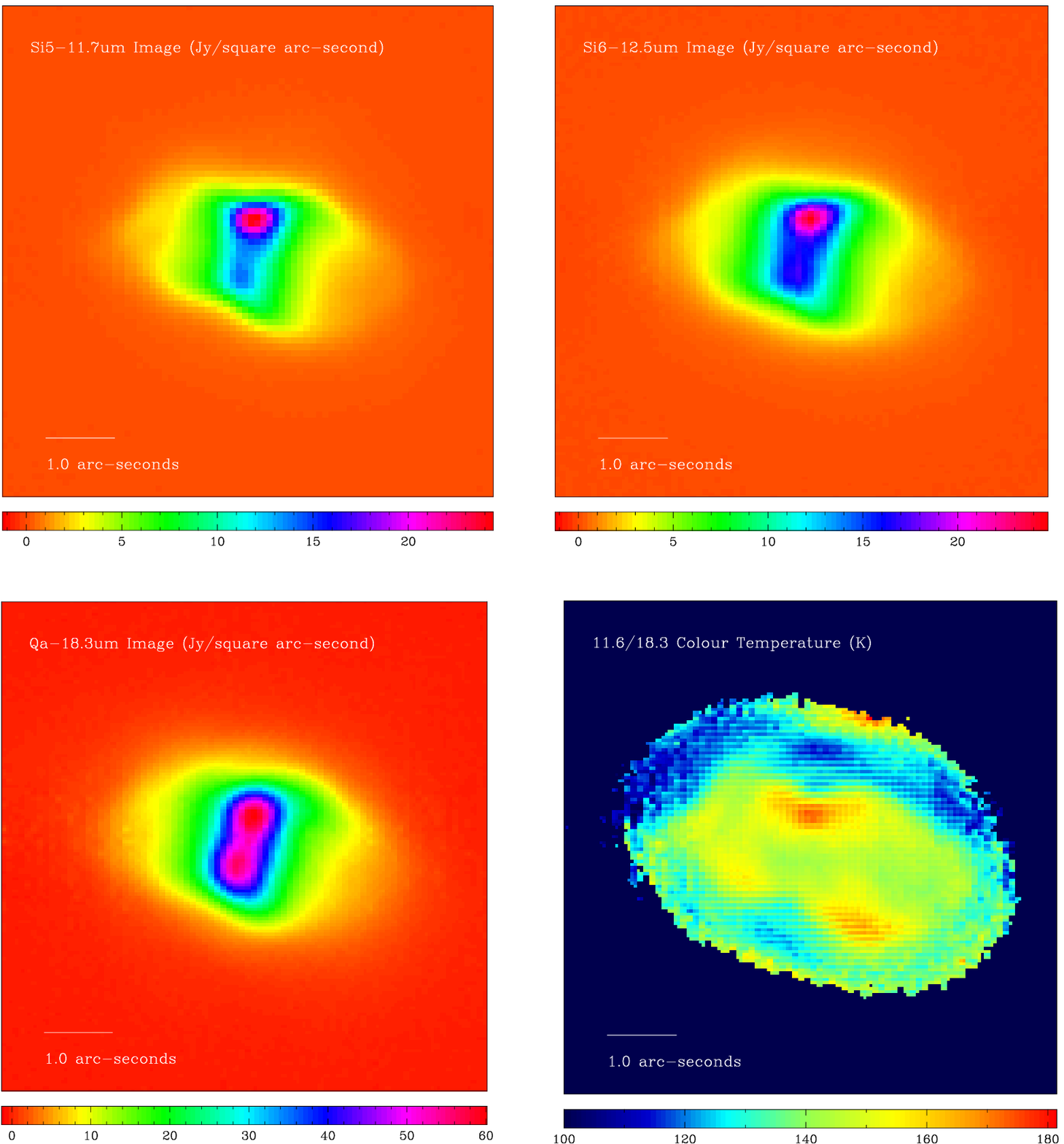} \figcaption[] {The flux-calibrated T-ReCS 
false-color images of IRAS 16594$-$4656.  The four panels give respectively 
the Si5 image (upper left), the Si6 image (upper right), the Qa image 
(lower left), and the color temperature map derived from the Si5 to 
Qa surface brightness ratio (lower right).  The T-ReCS images are 
given as surface brightnesses in units of Jy/square arc-second.  The 
color temperature values are given in K.  For each panel the bar gives 
the color mapping.  The image sections are all 7$\farcs$17 square.  All 
images have north up and east to the left.
\label{fig1}}

\clearpage

\epsscale{1.0} \plotone{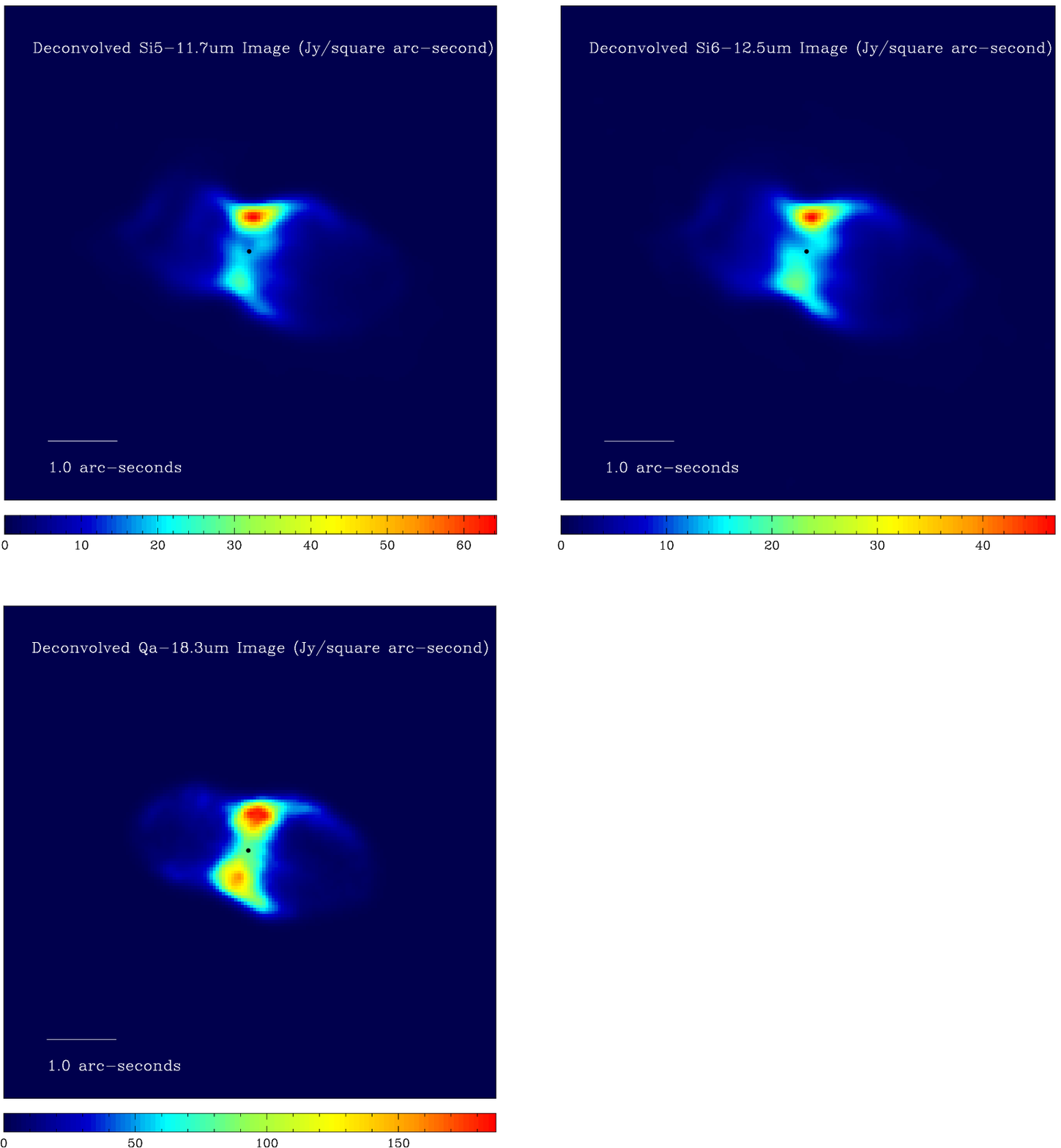} \figcaption[] {The deconvolved T-ReCS 
false-color images of IRAS 16594$-$4656.  The panels correspond to those 
in Figure \ref{fig1}.  In these images the estimated position of the star 
is marked by a small dot.\label{fig2}}

\clearpage

\epsscale{1.0} \plotone{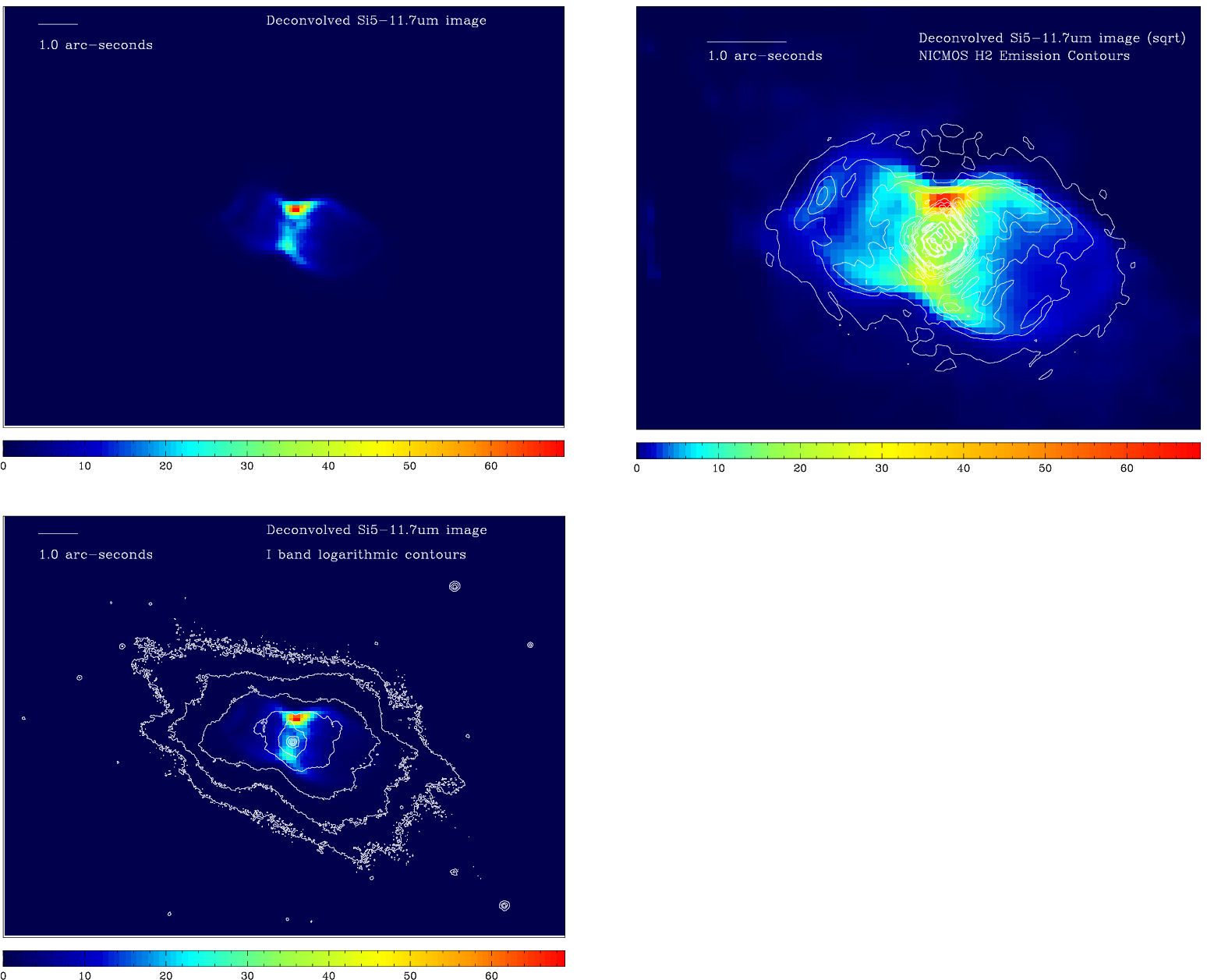} \figcaption[] {A comparison of the morphology 
of IRAS 16594$-$4656 observed at optical, near-infrared, and mid-infrared 
wavelengths.  The upper left panel shows the deconvolved Si5 image as in 
Figure 2.  The field of view is 14$\farcs$34 by 10$\farcs$75.  In the lower 
left panel the same figure is shown with logarithmically spaced contours from 
the {\it HST} I band image of \citep{su01} overlaid.  The upper right panel 
shows the Si5 filter image again, with square-root scaling to show the low 
level emission better; this panel is magnified by a factor of 2 to show the 
inner quadrant of the left panels.  Overlaid on the upper right T-ReCS image 
are linear contours from an H$_2$ image obtained with the {\it HST} NICMOS 
narrow-band 2.1212 $\mu$m emission-line filter \citep{hrivnak04}.
\label{fig3}}


\begin{thebibliography}{}

\bibitem[Cohen et al.(1999)]{cohen99}
    Cohen, M., et al. 1999, \aj, 117, 1864
\bibitem[Garc\'{i}a-Hern\'{a}ndez et al.~(2004)]{garcia04}
  Garc\'{i}a-Hern\'{a}ndez, D., {\it et al.}~2004, in {\it Asymmetrical 
Planetary Nebulae III}, eds.~M.~Meixner, J.~H.~Kastner, B.~Balick, \& 
N.~Soker, (ASP: San Francisco), 367
\bibitem[Garc\'{i}a-Lario et al.~(1999)]{garcia99}
  Garc\'{i}a-Lario, P., Manchado, A., Ulla, A., \& Manteiga, M. 1999, 
  \apj, 513, 941
\bibitem[Hrivnak, Kelly, \& Su (2004)]{hrivnak04}
  Hrivnak, B.~J., Kelly, D.~M., \& Su, K.~Y.~L.~2004, in {\it Asymmetrical 
Planetary Nebulae III}, eds.~M.~Meixner, J.~H.~Kastner, B.~Balick, \&
N.~Soker, (ASP: San Francisco), 175
\bibitem[Hrivnak, Kwok, \& Su (1999)]{hrivnak99}
  Hrivnak, B.~J., Kwok, S., \& Su, K.~Y.~L.~1999, \apj, 542, 849
\bibitem[Hrivnak, Kwok, \& Su (2001)]{hrivnak01}
  Hrivnak, B.~J., Kwok, S., \& Su, K.~Y.~L.~2001, \aj, 121, 2775
\bibitem[Hrivnak, Volk, \& Kwok (2000)]{hrivnak00}
  Hrivnak, B.~J., Volk, K., \& Kwok, S.~2000, \apj, 535, 275
\bibitem[Hrivnak et al.~(2006) in preparation]{hrivnak06}
  Hrivnak, B.~J., et al., 2006, paper in preparation
\bibitem[Kwok(1993)]{kwok93}
  Kwok, S.~1993, \araa, 31, 63
\bibitem[Rouleau \& Martin (1991)]{rouleau91}
  Rouleau, F., \& Martin, P.~G.~1991, ApJ, 377, 526
\bibitem[Santander-Garc\'{i}a et al.~(2004)]{santander04}
  Santander-Garc\'{i}a, M., Corradi, R.~L.~M., Balick, B., \& 
  Mampaiso, A.~2004, \aap, 426, 185
\bibitem[Su et al.~(2001)]{su01}
  Su, K.~Y.~L., Hrivnak, B.~J., \& Kwok, S.~2001, \aj, 122, 1585
\bibitem[Su et al.~(2001)]{su03}
  Su, K.~Y.~L., Hrivnak, B.~J., Kwok, S., \& Sahai, R.~2003, \aj, 126, 848
\bibitem[Ueta, Murakawa, \& Meixner (2005)]{ueta05}
  Ueta, T., Murakawa, K., \& Meixner, M.~2005, AJ, 129, 1625
\bibitem[Van de Steene \& Pottasch (1993)]{steene93}
  Van de Steene, G.~C. \& Pottasch, S.~R.~1993, \aap, 274, 895
\bibitem[Van de Steene, van Hoof, \& Wood (2000)]{steene00}
  Van de Steene, G.~C., van Hoof, P.~A.~M., \& Wood, P.~2000, \aap, 362, 984
\bibitem[Van de Steene \& van Hoof (2003)]{steene03}
  Van de Steene, G.~C. \& van Hoof, P.~2003, \aap, 406, 773
\bibitem[Volk \& Kwok (1989)]{volk89}
  Volk, K., \& Kwok, S.~1989, \apj, 342, 345
\bibitem[van Winckel (2003)]{vanwinckel03}
  van Winckel, H.~2003, \araa, 43, 391
\end{thebibliography}
\end{document}